\documentclass[a4paper,11pt]{article}
\usepackage{pos}
\usepackage{graphicx}
\usepackage{epsfig}

\title{A Fresh Look at the Chemical Potential on the Lattice}

\author*[a]{Rajiv V. Gavai}

\affiliation[a]{Indian Institute of Science Education \& Research Bhopal\\
Bhopal Bypass Road, Bhauri, Bhopal 462066, INDIA}

\emailAdd{gavai@tifr.res.in; rajiv@iiserb.ac.in}

\abstract{Lattice techniques are the most reliable ones to investigate the QCD
phase diagram in the temperature-baryon density (chemical potential) plane.
These techniques are, however, well-known to be saddled with a variety of
problems at nonzero density.  I address here the old question of placing the
baryonic (quark) chemical potential on the lattice and point out its important
consequences for the current and future experimental searches of the QCD
critical point.}

\FullConference{%
 The 38th International Symposium on Lattice Field Theory, LATTICE2021
  26th-30th July, 2021
  Zoom/Gather@Massachusetts Institute of Technology
}

\begin{document}
\maketitle

\section{Introduction}

\begin{figure}[htb]
\begin{center}
\includegraphics[scale=0.85]{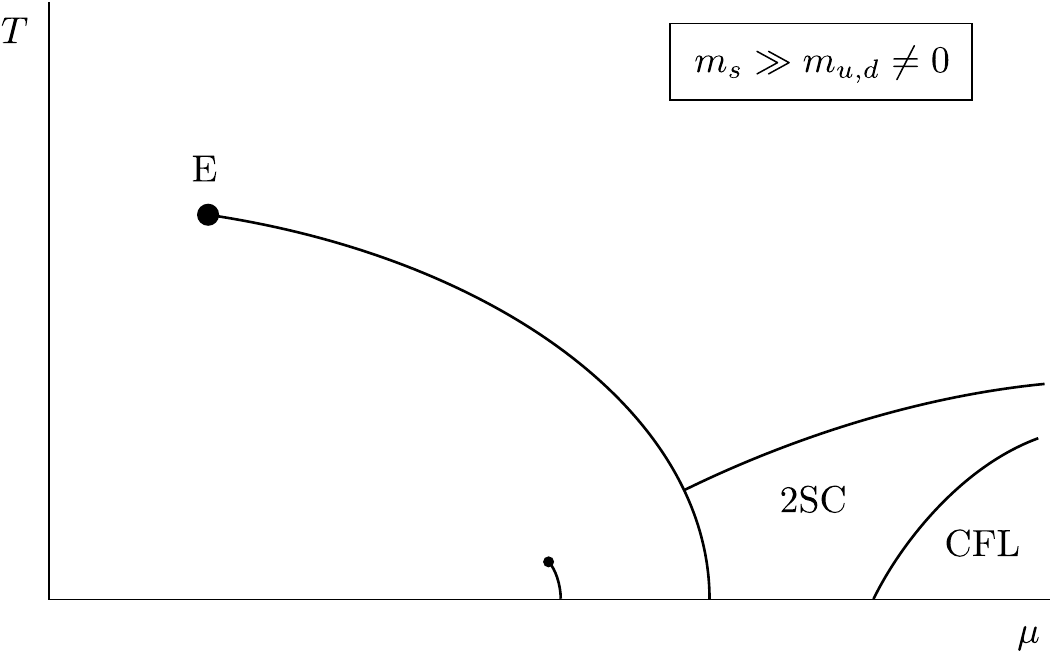}
\caption{A typical QCD phase diagram in the $T$-$\mu_B$ plane, showing the QCD
critical point for $m_u = m_d >> m_s$. Taken from Ref.~\cite{Raja}.}
\label{qcdc}
\end{center}
\end{figure}

Quantum chromodynamics (QCD) is now widely accepted as the theory of strong
interactions. However, simple phenomenological models were in vogue much before
its heralding. In view of the strong non-perturbative nature of the hadronic
world these models continue to be employed for qualitative, and at times even
quantitative, understanding of the underlying physics.  The behaviour of
strongly interacting matter under extreme temperatures and/or densities is one
such area.  Based on chiral symmetries of QCD and the corresponding Nambu-Jona
Lasinio type of models the expected phase diagram for 2 light and one moderately
heavy quark has emerged \cite{Raja}, as displayed in Figure \ref{qcdc}.  One of
its fundamental aspect is the critical point in $T$-$\mu_B$ plane, where $\mu_B$
is the baryon chemical potential which governs the net baryon density. 

A lot of theoretical and experimental attention has been focused on locating the
QCD critical point, and in devising its characteristic signatures.   In
particular, it was suggested that higher moments of fluctuations are enhanced
near a critical point and a non-monotonic behaviour in them as a function of
colliding energy could be a signal for criticality \cite{HigMom}. Inspired by
lattice QCD results, employing baryonic higher order fluctuations was advocated
\cite{ggplb}.  Exciting experimental results \cite{star14} from the STAR
collaboration on net proton fluctuations seemed intriguingly like the
theoretical expectations \cite{ggplb}.  Sixth order baryonic fluctuations have
been advocated as a possible tool to establish the expected $O(4)$-criticality
for the two-flavour QCD at nonzero temperatures with $\mu_B=0$ \cite{o4crit}.
Finally, various methods to determine the location of the critical point using
lattice techniques, specifically the Taylor series method, need higher order
baryonic fluctuations.  It thus appears worthwhile to examine the higher order
quark number susceptibilities (QNS) carefully which we set out to do here.

\section{The $\mu \ne 0$ problems}

It is well-known that the fermionic determinant becomes complex for nonzero
quark chemical potential.  This leads to the well-known sign/phase problem due
to the lack of positivity of the measure of the corresponding path integral.  
Several approaches have been proposed, many of which rely eventually on using
the higher derivatives of determinant related to the QNS.  To that extent our
arguments below would apply to many of them, though for definiteness we will
focus on the Taylor series expansion method where successive higher order terms
are needed to improve it systematically. 

Mostly staggered fermions are used in the investigations of QCD phase diagram.
These do not have a well defined flavour number or the $U_A(1)$ symmetry on the
lattice.  While these may be attained in the continuum limit on large lattices,
it is debatable as to how well the present days lattice simulations perform from
this perspective.   Model considerations suggest that two light quark flavour
QCD with an anamoly contribution which decreases slowly with temperature at zero
chemical potential leads to a QCD critical point in the temperature-baryon
density phase diagram.  Higher number of light flavours or a sharp drop in the
anamoly contribution may lead only to a first order transition line in the phase
diagram.   Domain wall quarks or the overlap quarks have the same symmetries as
the continuum quarks but do not have a unique local quark current, making
it difficult to define the chemical potential for them.  There have been
proposals in the literature with their associated problems.  We argue that the
problem of definition of chemical potential generically affects all types of
fermions, and that demanding universality of the QNS leads us to a definition
valid for all types.

For simplicity, we shall consider the simplest case of na{\i}ve fermions as the
arguments below can be easily extended to all others with similar consequences.
Recall that the na{\i}vely discretized fermionic action is
$$ S^F = \sum_{x,x'} \bar\psi(x){\left[\sum^4_{\mu=1}D^{\mu}(x,x') 
+ m a \delta_{x,x'} \right]}~\psi(x'), $$
where
$$ D^\mu(x,x') = {\frac {1}{2}} \gamma^\mu \left[U^\mu_x\delta_{x,x'-\hat\mu} -
U^{\mu\dagger}_{x'}\delta_{x,x'+\hat\mu}\right].
$$
It clearly has a $U(1)$ phase rotation symmetry. Under $\psi' = \exp{(i \alpha)}
\psi$ and $ \bar \psi' = \bar \psi \exp{(-i \alpha)}$, with $0 \le \alpha < 2
\pi$ and constant, it remains invariant.  As usual, one can easily follow the
canonical method to write the corresponding current conservation equation,
$\sum_\mu \Delta_\mu J^{lat}_\mu =0$, and obtain from it the conserved charge in
the naturally point-split form.  Switching on the gauge fields, $U^\mu(x)$, the
conserved quark/baryon charge in QCD is $N= \sum_x \bar \psi(x)
\gamma^4[U^4_x\psi(x+\hat 4) + U^{4 \dagger}_{x-\hat 4} \psi(x-\hat 4)]/2$.
Adding the baryonic chemical potential to the action therefore  amounts to
weights $f(a \mu) = 1 + a \mu$ and $g(a \mu) = 1 - a \mu$ to forward and
backward time links respectively in the full lattice QCD action.  Even for the
free case, $\forall U^\mu(x) = I $, this action leads to $\mu$-dependent
$a^{-2}$ divergences in the $a \to 0$ limit in energy density $\epsilon$, and 
quark number density $n$. In particular, one has  

\begin{eqnarray}
\epsilon &=& {\bf c_0a^{-4} + c_1\mu^2a^{-2}} + c_3\mu^4 + c_4\mu^2T^2 + c_5T^4 \\
\nonumber n &=& {\bf d_0a^{-3} + d_1\mu a^{-2}} + d_3\mu^3 + d_4\mu T^2 + d_5T^3. 
\label{ElNl}
\end{eqnarray} 

Subtracting off the vacuum contribution at $T = 0 = \mu$, eliminates the leading
divergence in each case but the $\mu$-dependent divergence persists.  This has
been, of course, known since long, and proposals to eliminate these divergences
also exist.   Hasenfratz and Karsch \cite{hk} as well as Kogut et al.
\cite{kogut}  proposed to modify the weights  $f$ and $g$ to $\exp (\pm a \mu)$
to obtain finite results while simultaneously Bili\'c-Gavai \cite{bilgav} showed
$( 1 \pm a \mu)/ \sqrt ( 1- a^2 \mu^2)$ also lead to finite results.
Indeed, in general {\em any} set of functions $f$, $g$,
satisfying $ f(a \mu) \cdot g(a \mu) =1$ with $f(0) = f'(0) =1$ suffices to
eliminate the $\mu$-dependent divergence \cite{gavai}.

It is worth emphasizing that the analytical proof of elimination of divergences
was provided {\em only} for free quarks \cite{hk,bilgav, gavai}, and thus 
also for perturbation theory.  Numerical computations had to be
performed to show that it worked for the non-perturbative interacting case as
well \cite{ggq}.

\begin{figure}[htb]
\begin{center}
\includegraphics[scale=0.515]{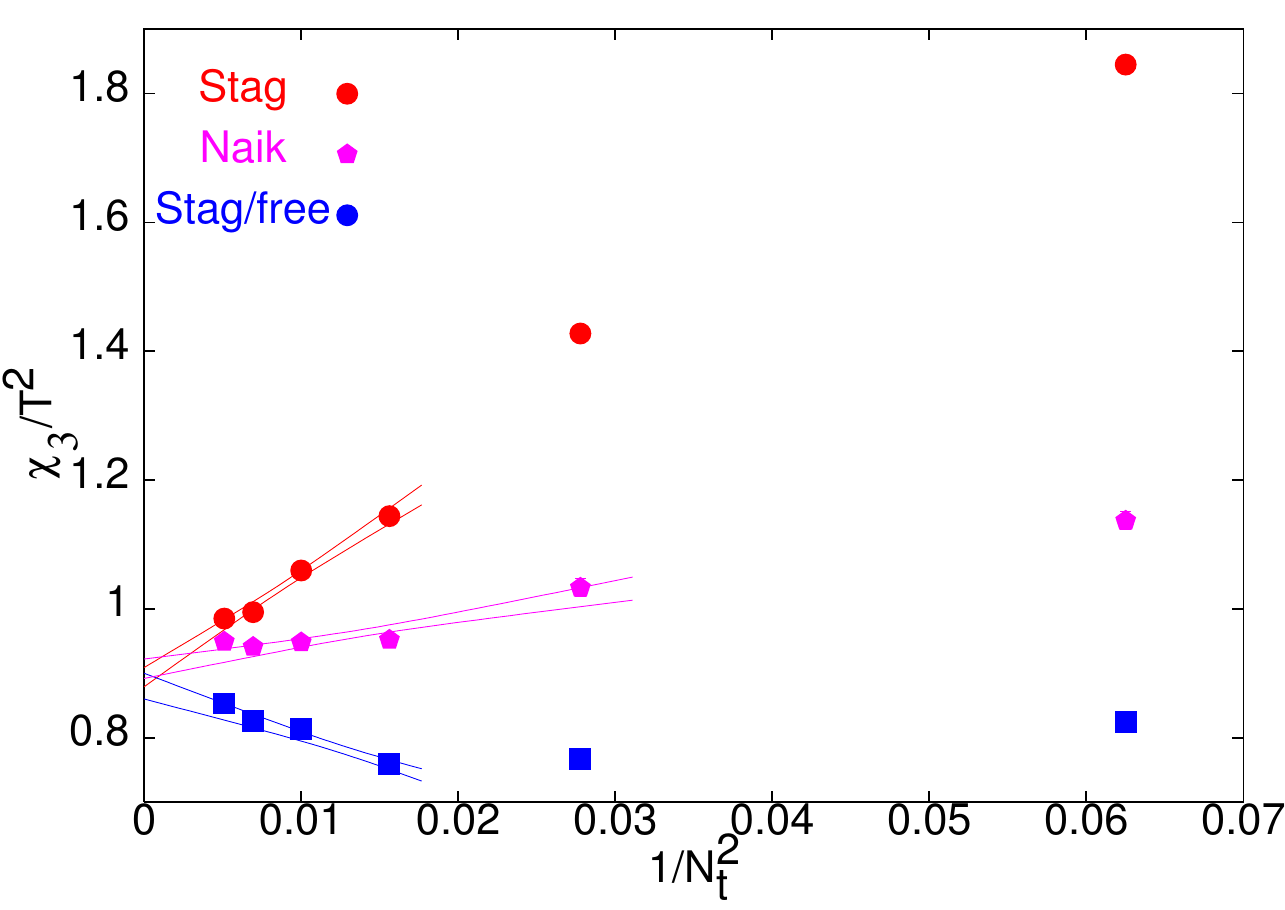}
\caption{Continuum limit for quark number susceptibilities with different
actions. A linear behaviour of the data and convergence to a unique continuum 
limit indicates the absence of any divergence. Taken from Ref.~\cite{ggq}.}
\label{univ}
\end{center}
\end{figure}
Figure \ref{univ} displays the results of \cite{ggq} for the continuum limit of
quark number susceptibilities with different actions computed for $\mu=0$ at a
constant temperature of 2$T_c$.  Since $\chi \sim \partial n / \partial \mu$, it
too has a quadratic divergence which naturally is $\mu$-independent.
Furthermore, at fixed temperature $T^{-1} = N_t \cdot a$, the continuum limit of
$a \to 0$ is equivalent to $N_t \to \infty$.  The data in Figure \ref{univ}
exhibit a linear behaviour and converge to a unique limit to the expected
result at that temperature. Since a rising curve to infinity as $ 1/N_t \to 0$
would be the signal of a divergence, the absence of divergence is demonstrated
in a non-perturbative manner as well for all $ f(a \mu) \cdot g(a \mu) =1$
choices of the weight factors as the $\chi$ is the same for all of them.

One may naturally wonder why this type of divergence was not encountered in the
continuum theories.  Specifically, one may attribute it to the latticization.  A
question then arises as to why, and how, does lattice introduce this divergence.
An argument was provided \cite{hk} in the early work, stating that the
divergence arises on the lattice due to the lack of a "formal" gauge symmetry.
In the continuum theory, the $\mu$ term in the action appears as if it is a
4$^{th}$ component of a constant gauge field. The simple na{\i}ve choice for the
weights does not respect such formal global gauge invariance on the lattice
while all the forms above where divergences are eliminated do have this formal 
gauge symmetry.
\begin{equation}
 f(a \mu) \cdot g(a \mu) =1  \Leftrightarrow F(a \mu) = \exp (\ln f(a \mu)),
 G(a \mu) = \exp (-\ln f(a \mu)).
\end{equation}

It is important to note that even this formal analogy with the global
gauge transformation strictly speaking needs an imaginary chemical potential and
is thus invalid for the physically interesting case of the real chemical
potential.  Not surprisingly, it turns out to be {\bf wrong} as well: 
Latticization does not introduce divergence. It merely assists in spotting 
what exists in the continuum already \cite{GaSh}.  Techniques such as contour
integrals in the continuum have an inbuilt way to eliminate these divergences,
either by subtraction or by a suitable regulator-based prescription.  It has
been demonstrated \cite{GaSh} that a similar subtraction scheme can be used on
the lattice to remove the divergent contributions, leading to finite results in
the $a \to 0$ limit also for the action with the linear chemical potential.

\section{Universal Or Unique?}

As discussed above, there are three different lattice action for finite density
QCD, including the na{\i}ve one, for all of which divergences can be eliminated
from the continuum limit result.  A priori, the multiplicity of lattice actions
corresponding to the same continuum QCD action is a familiar phenomenon.
Indeed, the entire program of improvement of action, which aims to obtain better
results even for not so small $a$, relies on the concept of {\em universality}
which assures the same universal results for all physical quantities in the $a
\to 0$ limit.  As we will argue below, there are subtleties in the case of
finite density actions which raise the question posed in the heading of this
section.  

Recall that the three lattice actions for $\mu \ne 0$ are characterised by the
three different weight factors $f(\mu a)[g(\mu a)]$ for all the forward
[backward] time-like links.  Denoting them by $L$, $E$ and $S$ for their linear,
exponential and square root forms, the corresponding weights are given by  

\begin{eqnarray}
f_L (\mu a) &=& 1 + \mu a~,  \qquad\qquad\qquad\qquad g_L(\mu a) = 1 -\mu a\\ \nonumber
f_E (\mu a) &=& \exp(\mu a)~, \qquad\qquad\qquad\qquad g_E(\mu a) = \exp (-\mu a)\\ \nonumber
f_S (\mu a) &=& (1 + \mu a)/\sqrt{1-\mu^2 a^2}~, \qquad g_S(\mu a) = (1 -\mu
a)/\sqrt{1-\mu^2 a^2}~.~
\label{3acts}
\end{eqnarray}

Considering the classical continuum limit of the actions incorporating these
weight factors, it is easy to verify that all of them have the same form at
${\cal O}(\mu a)$.  Indeed, these actions all differ in terms starting ${\cal O}
(\mu^2 a^2)$. Universality of physical results is assured if all these terms are
irrelevant. Alternatively, if these actions produce different results, one ought
to worry whether these terms are indeed irrelevant, and can thus be legitimately
be added to the action at all. Assuming that they are irrelevant leads to the
following paradox. It has been shown that the $L$-action has divergences
mentioned earlier, which have to be subtracted by hand while the other two do
not have them in an apparent violation of universality.  Furthermore, since the
divergences do exist in the continuum theory, as demonstrated in Ref.
\cite{GaSh}, one wonders how/whether they will reappear for the actions with
$f_E$ or $f_S$ as $a \to 0$.  As a next step to assess how universal the
physical results from these actions are let us examine \cite{rvgchem} the quark
number susceptibility at $\mu=0$ in some details in order to understand this
difference.  It is given by

\begin{equation}
   \chi = \frac{1}{N_t N_s^3 a^2}
\left[\left\langle\left({\rm Tr} M^{-1} M'\right)^2
                     \right\rangle +
         \left\langle{\rm Tr}
             \left(M^{-1} M'' - M^{-1}M'M^{-1}M'\right)
                  \right\rangle\right],
\end{equation}
Here $M$ is the quark matrix (inverse propagator) and $M'$($M'')$ is its
first(second) derivative with respect to $\mu$. We shall consider $\chi(\mu=0)$,
and therefore $M$, $M'$ \& $M''$ above are evaluated at $\mu=0$. None of
the terms depends on $a$ explicitly [$S_F(a\mu) = \bar \psi M(a \mu) \psi$].
All the terms inside the bracket are dimensionless, with the $a^2$ factor in the
denominator supplying the dimension of $\chi$. In the continuum limit of $a 
\to 0$, {\em all} the terms must vanish at least as $a^2$ for a 
nontrivial continuum limit of $\chi$.  Of course, if any term vanishes as $a^n$,
$n \ge 3$, it will be irrelevant in the continuum limit while $n \le 1$ would
lead to a divergence.  While comparing the three $\chi$'s, it should be noted
that $M'$ is the same for all $f$'s while $M'' = 0$ for $f_L$ and nonzero for
others.  Thus the presence of quadratic divergence for the linear form can be
attributed to the vanishing of its $M''$ term, implying that the other two terms
together have $a^0$ as the leading behaviour.  The exponential and square root
form have the same nonzero $M''$ at $\mu=0$, coming from the $\mu^2 a^2$ terms
in them.  Its presence precisely cancels the $a^0$ piece of the other two terms
leading to a divergence-free result for $\chi$, as has been demonstrated for
both the free theory(analytically) \cite{hk,bilgav,gavai} and the quenched QCD
(numerically) \cite{ggq}.  Therefore the $M''$ term must also have $a^0$, and it
is ruled out that it vanishes faster than $a^2$ for any action.  Hence, the
difference in the results remains even in the continuum limit, in spite of $M''$
arising from an apparent {\bf irrelevant} term in the action, namely ${\cal
O}(\mu^2a^2)$ term. 

It is clear that this problem of {\em non-universal} results worsens as
one computes higher order fluctuations.  Each successive order will have terms
with the corresponding higher derivative of $M$, which in turn means higher
derivatives of $f$ (since $g$ = 1/$f$ is assumed to cancel the divergences).
While $f_E''(0) = f''_S(0)$, it is easy to show that  $f_E'''(0) = f''''_E(0)= 
1$ but $f_S'''(0) =3$.  Moreover, $f''''_S(0)=9$ \cite{rvgchem}.
Indeed, $f_E^n(0) \ne f^n_S(0)$ for all $n>2$.  Let us recall that the
third and fourth order baryon number susceptibilities, which are themselves
related to various quark number susceptibilities up to the corresponding order,
are crucially used in locating \cite{rvgcnph} the QCD critical point, both 
theoretically and experimentally.  In fact, one hopes to improve upon the
current results by incorporating as many higher orders as possible.  
Furthermore, the sixth order baryonic fluctuations have been advocated as 
signature for the $O(4)$ criticality \cite{o4crit} which too will need
up to sixth order of the quark number susceptibilities.

As we argued above, the theoretical results for all these fluctuations 
{\em depend} on which $f$ one uses irrespective of the choice to eliminate 
the divergence by the $f \cdot g =1$ condition or by subtraction. One has the
embarrassing situation of having as many predictions for physically measurable
quantities as the number of $f$ !   Thus the desire to eliminate the divergences
by the $f \cdot g =1$ condition does not appear to be in accord with
universality.

Trying to understand the reasons beneath this violation of universality, we note
that $\mu$, being dimensionful, one has to ensure that it has a fixed value in
physical units as $a \to 0$. Recall that in order to achieve the continuum limit
at a fixed temperature $T =1/N_t a$ in physical units one needs to take $a \to
0$ {\em along with} $N_t \to \infty$. Thus not only $a$ but $Ta=1/N_t$ are two
separate scales which need to be tuned appropriately.   Similarly, one will have
to treat $a$ and $\mu a$ as separate scales which will have to be approach zero
by themselves such that $\mu$ is held the same in physical units.  At the purely
classical level, where one examines the approach of the lattice action to that
in the continuum, the additional ${\cal O}(\mu^2a^2)$ term can be easily seen to
vanish {\bf iff} $a \to 0$ limit is the same as $\mu a \to 0$.  At the
quantum expectation level, on the other hand, we expect $\mu/T$ to be held fixed
as $a \to 0$, making $\mu a \to 0$ a separate limit.
The $n^{th}$ derivatives of $P$ will be, by definition, sensitive to  the
corresponding $\mu^na^n$ terms in $P$ for $n \ge 2$,  exposing their 
relevant nature in the corresponding susceptibilities, and consequently to
pressure.  It needs to be emphasized that while the entire effort to modify the
$f_L$ and $g_L$ from their natural linear prescription, which can be derived 
from an underlying conserved current on the lattice, was motivated by the
question of divergences in eq,(\ref{ElNl}), the non-universality pointed out
above has nothing to do with the diverging terms.  It is primarily related to
the question of adding term to all orders in $\mu a$ with the assumption/hope
that universality will make their contribution irrelevant.  The elimination of
divergence is, in fact, a proof that it is not so, {\em i.e.}, $\mu a$ being a
separate scale which needs to be tuned in the continuum limit these terms are
relevant. We advocate that it is more advisable to use the {\bf unique} $f_L$ 
and $g_L$ in lattice computations than to sacrifice universality.

Changing from the linear form to either the exponential or the square root form
leads to a further problems. The actions with either $f_E$ or with $f_S$ has no
conserved charge anymore \cite{rvgchem}. Following the canonical method to
derive current conservation for actions with or without chemical potential terms
in the continuum, or on the lattice, it is easy to prove this.  Both in the
continuum and for the $f_L$,$g_L$ lattice action, the {\em same} current
conservation equation results\cite{rvgchem} for $\mu \ne 0$ as for $\mu =0$, and
thus the conserved charge is unaffected by $\mu \ne 0$ terms, as it must.  On
the other hand, for $f_E$ and $f_S$-actions with $\mu \ne 0$ one gets a
conserved charge that itself depends on $\mu$ !  As a direct consequence, $Z
\ne \sum_n z^n Z^C_n $ on the lattice for these two actions. This is possible
only if the $a \to 0$ and $\mu a \to 0$ limits are identical for {\em all
$\mu$}.  This implies that one cannot define an {\em exact} canonical partition
function on lattice from the $Z$ defined this way for the actions employing the
exponential or the square root forms.  Only for the linear $\mu$-case one has an
exactly conserved charge on the lattice, and thus an {\em exact} canonical
partition function $Z = \sum_n z^n Z^C_n $ on the lattice.

\begin{figure}[htb]
\begin{center}
\includegraphics[scale=0.35]{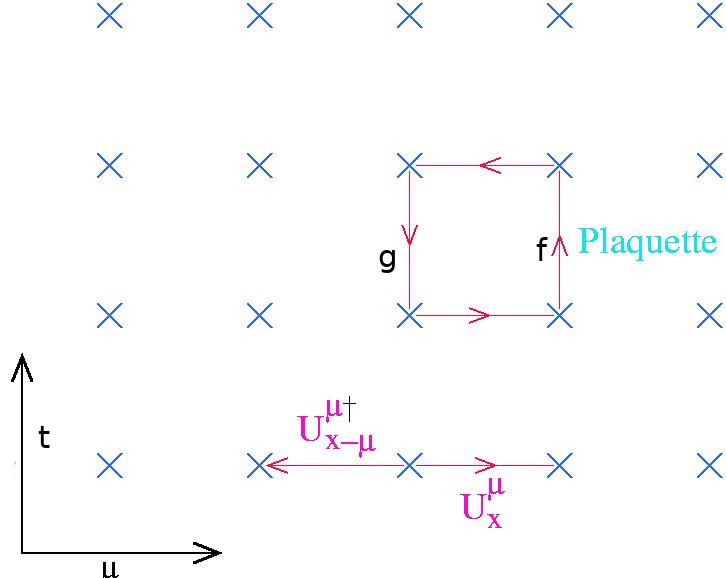}
\caption{Illustration of the weight factors in the $t$-$\mu$ plane ($\mu=$1,2,3).} 
\label{latmu}
\end{center}
\vskip -0.2 cm
\end{figure}

The $f \cdot g =1$ condition introduces differences in the $\mu$-dependence of
the partition function $Z$ which again seem to survive the $a \to 0$ continuum
limit.  Given the fact that only timelike links acquire a nontrivial weight
factor at $\mu \ne 0$, it is obvious that any $\mu$-dependence for $Z$ arises
solely due to loops with time-like links, and hence is $\propto (f \cdot g)^l $,
where $l$ is the number of forward or backward links in a Wilson loop.  Figure
\ref{latmu} illustrates this for $l=1$, the timelike plaquette. Since  $ f_L
\cdot g_L = 1 -\mu^2 a^2$, quark loops of {\bf all} sizes and types contribute
to the $\mu a$ dependence for the $Z$.  This is similar to the expectation in
the continuum as well where time-like Wilson loops of {\bf all} sizes would
contribute.  However, only quark loops winding around the $T$ direction
contribute to $\mu$ dependence for the other two cases since $f \cdot g = 1 $.
The $\mu$-dependence thus arises from only a topologically distinct class of
Wilson lines, raising  the same universality violation issue again.  Of course,
only explicit calculations may show what is the fate of the topologically
trivial Wilson loops of all sizes and whether they somehow add up to a
$\mu$-independent expression.  At least, this apparent paradox suggests that it
is crucial to check if universality is obeyed for all three actions perhaps by
considering other theories which do not have the sign problem for nonzero $\mu$.

\section{Summary}

Prescriptions to modify weights for nonzero chemical potential in the lattice 
actions which satisfy $f \cdot g =1$,  where $f(\mu a)$[$g(\mu a)$] is the 
weight factor for a forward [backward] timelike gauge link, such as the popular
exponential form,  violate universality for quark number susceptibilities.  As
their order increases, further non-universal results arise depending on the
exponential\cite{hk,kogut} or the square root form \cite{bilgav}. Employing them
leads to $f$-dependent results even in the continuum limit for fluctuations
measurable in, and of great interest to, heavy ion collision experiments.

$\mu$-dependent divergences, for elimination of which the condition was
invented, are not unique to lattice regularization. Indeed, lattice only
reproduces faithfully what exists in the continuum field theory. Following the
example of continuum field theory, one can subtract the free theory 
divergences by hand, and this process has been shown to suffice even 
nonperturbatively \cite{GaSh}. The linear form preserves the quark/baryonic
current conservation on the lattice, and retains the same conserved quark/baryon
number for all $\mu$.  On the other hand, the conserved charge depends on $\mu
a$ for the other two. As a result the canonical partition function can
meaningfully be defined on the lattice exactly only for the linear form.

\end{document}